\documentclass[graphicx]{nato}
\usepackage{sprrefn}
\usepackage{color}

\begin{document}

\begin{opening}
\title{Superconducting qubits II: Decoherence}
%\subtitle{including microscopic aspects of superconducting qubits}
\author{F.K. Wilhelm\footnote{present address: Physics Department and Insitute
for Quantum Computing, University of Waterloo, Waterloo, Ontario N2L 3G1, Canada; {\tt fwilhelm@iqc.ca}}}
\author{M.J. Storcz}
\author{U. Hartmann}
\institute{Department Physik, Center for Nanoscience, and Arnold Sommerfeld Center for theoretical physics, Ludwig-Maximilians-Universit\"at, 80333 M\"unchen, Germany}
\author{Michael R. Geller\thanks{mgeller@uga.edu}}
\institute{Department of Physics, University of Georgia, Athens,  Georgia 30602, USA} 
\runningtitle{Superconducting qubits II: Decoherence}
\runningauthor{F.K. Wilhelm, M.J. Storcz, U. Hartmann, and M. Geller}
\begin{abstract}
This is an introduction to elementary decoherence theory as it is typically applied to superconducting qubits. 
\end{abstract}
\abbreviations{\abbrev{SQUID}{superconducting quantum interference device}; \abbrev{qubit}{quantum bit}; \abbrev{TSS}{two state system}}
\end{opening}
\tableofcontents

\section{Introduction}

The transition from quantum to classical physics, now known as decoherence,  has intrigued physicists since the formulation of quantum mechanics \cite{Joos96,Leggett02,Peres93,Feynman63,Zurek93}. It has been put into the poignant Schr\"odinger cat paradox \cite{Schroedinger35} and was considered an open fundamental question for a long time. 

In this chapter, we study the theory of decoherence as it is applied to superconducting qubits. The foundations of the methodology used are rather general results of quantum statistical 
physics and resemble those applied to 
chemical physics, nuclear magnetic resonance, optics, and other condensed matter systems 
\cite{Weiss99}. 
All these realizations introduce their subtleties --- typical couplings, temperatures, 
properties of the correlation functions. We will in the following largely stick to effective spin notation in
order to emphasize this universality, still taking most of the examples from superconducting 
decoherence. This paper is based on lectures 2 and 3 of the NATO-ASI on ``Manipulating
quantum coherence in superconductors and semiconductors" in Cluj-Napoca, Romania, 2005. It is 
{\em not} intended to be a review summarizing the main papers in the field. Rather, it is an (almost) self-contained introduction to some of the relevant techniques, aimed to be accessible to researchers and 
graduate students with a knowledge of quantum mechanics \cite{Cohen92} and statistical physics \cite{Landau84} on the level of a 
first graduate course. So much of the material here is not new and most certainly known to more
experienced researchers, however, we felt a lack of a single reference which allows newcomers to get started without excessive overhead. References have largely been chosen for the aid they 
provide in learning and teaching the subject, rather than importance and achievement. 

\subsection{Basic notions of decoherence}

The mechanisms of decoherence are usually related to those of energy dissipation. In particular, 
decoherence is irreversible. If we take as an example a pure superposition state
\begin{equation}
|\psi\rangle =(|0\rangle+|1\rangle)/\sqrt{2}\quad \rho_{\rm pure}=|\psi\rangle\langle\psiÊ|
=\frac{1}{2}\left(\matrix{1&1\cr 1&1\cr}\right)
\end{equation}
and compare it to the corresponding classical mixture leading to the same expectation value 
of $\sigma_z$
\begin{equation}
\rho_{\rm mix}=\frac{1}{2}\left(\matrix{1&0\cr 0&1\cr}\right)
\end{equation}
we can see that the von-Neumann entropy $\rho=-k_B{\rm Tr}\left[\rho\log\rho\right]$ rises from $S_{\rm pure}=0$
to $S_{\rm mix}=k_B\ln 2$. Hence, decoherence taking $\rho_{\rm pure}$ to $\rho_{\rm mix}$ creates entropy and is irreversible. 

Quantum mechanics, on the other hand, is always reversible. It can be shown, that any {\em isolated}
quantum system is described by the Liouville von-Neumann equation 
\begin{equation}
i\hbar\dot\rho=[H,\rho]
\label{eq:Liouville}
\end{equation}
which
conserves entropy. Indeed, also the CPT theorem of relativistic quantum mechanics \cite{Sakurai67} states, that for each quantum system it is possible to find a counterpart
(with inversed parity and charge) whose time arrow runs backwards. The apparent contradiction between microreversibility --- reversibility of the laws of quantum physics described by Schr\"odinger's
equation --- and macro-irreversibility is a problem at the foundation of statistical thermodynamics. We also remark that the Lagrangian formalism \cite{Landau82} which was used as the starting point in the previous chapter of this book \cite{Geller06} does not even accomodate friction on a classical level without artificial and in general non-quantizable additions. 

\subsubsection{Heat baths and quantum Brownian  motion \label{qbm}}

The standard way out of this dilemma is to introduce a continuum of additional degrees of freedom
acting as a heat bath for the quantum system under consideration \cite{Feynman63,Caldeira81,Caldeira83}. The complete system is
fully quantum-coherent and can be described by equation \ref{eq:Liouville}. However, the heat bath contains  unobserved degrees of freedom which 
have to be integrated out to obtain the {\em reduced  system}; the reduced system is the original 
quantum system which does not contain the bath explicitly, but whose dynamics are influenced by the bath. The dynamics of the reduced system 
now show both {\em dissipation} (energy exchange with the heat bath) and {\em decoherence} (loss of quantum information to the heat bath). Another view on this is that any finite combined quantum system shows dynamics which are periodic in time. The typical periods are given by the inverse level splittings of the system. Thus, a continuous heat bath shows periodicity and reversibility only on an infinite, physically unobservable time scale. 

A standard example, taken from Ref. \cite{Ingold98}, of irreversibility in both classical and quantum mechanics is (quantum) Brownian motion (QBM), which we will now describe in the one-dimensional case. The underlying Hamiltonian of a single particle in an oscillator bath has the general structure
\begin{equation}
H=H_s+H_{sb}+H_b+H_c.
\end{equation}
Here, the system Hamiltonian $H_s$ describes an undamped particle of mass $M$ in a scalar potential, $H_s=\frac{P^2}{2M}+V(q)$. $H_b$ describes a bath of harmonic oscillators, 
$H_i=\sum_i \left(\frac{p_i^2}{2m_i}+\frac{1}{2}m_i\omega_i^2x_i^2\right)$. The coupling between these
two components is bilinear, $H_{sb}=-q\sum_i c_i x_i$. If this were all, the effective potential seen by the 
particle would be altered even on the classical level, as will become more obvious later on. Thus, 
we have to add a counter term which does not act on the bath, $H_c=q^2\sum_i \frac{c_i^2}{2m_i\omega_i^2}$. Adding this counterterm gives the Hamiltonian the following intuitive form
\begin{equation}
H=\frac{P^2}{2M}+V(q)+\sum_i \left(\frac{p_i^2}{2m_i}+\frac{1}{2}m_i\omega_i^2\left(x_i-\frac{c_i}{m_i\omega_i^2} q\right)^2\right)
\label{eq:hamiltonian_qbm}
\end{equation}
indicating that the bath oscillators can be viewed as attached to the particle by springs. Here, we have introduced sets of new parameters, $c_i$, $\omega_i$, and $m_i$ which need to be adjusted to the system of interest. This aspect will be discussed later on. We treat this system now
using the Heisenberg equation of motion 
\begin{equation}
i\hbar\dot{O}(t)=\left[O(t),H\right]
\end{equation}
for the operators $q$, $P$, $x_i$, and $p_i$, which (as a mathematical consequence of the correspondence principle) coincide with the classical equations of motion. The bath oscillators see the qubit acting as an external force
\begin{equation}
\ddot{x_i}+\omega_i^2x_i=\frac{c_i}{m_i}q(t).
\end{equation}
This equation of motion can be solved by variation of constants, which can be found in 
textbooks on differential equations such as \cite{Zill00}
\begin{equation}
x_i(t)=x_i(0)\cos\omega_i t+\frac{p_i(0)}{m_i\omega_i}\sin\omega_i t+ \frac{c_i^2}{m_i\omega_i^2}\int_0^t dt^\prime\sin\omega_i (t-t^\prime)q(t^\prime)
\label{eq:bath_osci_formal}
\end{equation}
Analogously, we find the equation of motion for the particle
\begin{equation}
\ddot{q}=-\frac{\partial V}{\partial q}-\sum \frac{c_i}{m_i}x_i -q\sum\frac{c_i^2}{m_i\omega_i^2}.
\label{eq:particle_formal}
\end{equation}
Substituting eq. \ref{eq:bath_osci_formal} into eq.\ \ref{eq:particle_formal} eliminates the bath coordinates up to the initial condition
\begin{eqnarray}
M\ddot{q}&=&-\frac{\partial V}{\partial q}-\sum_i \frac{c_i^2}{m_i\omega_i}\int_0^t dt^\prime
\sin\omega_i(t-t^\prime)q(t^\prime)\nonumber\\
&&+\sum_i c_i \left(x_i(0)\cos\omega_i t+\frac{p_i(0)}{m_i\omega_i} 
\sin\omega_i t\right)-q\sum_i \frac{c_i^2}{m_i\omega_i^2}.
\end{eqnarray}
We now integrate by parts and get a convolution of the velocity plus boundary terms, one of which shifts 
the origin of the initial position, the other cancels the counterterm (indicating, that without the 
counterterm we would obtain a potential renormalization). The result has the compact form 
\begin{equation}
M\ddot{q}+\frac{\partial V}{\partial q}+\int_0^t dt^\prime \gamma(t-t^\prime) \dot{q}(t^\prime)
=\xi(t).
\label{eq:Langevin}
\end{equation}
This structure is identified as a Langevin equation with memory friction.
If interpreted classically, this is the equation of motion of a Brownian particle - a light particle in
a fluctuating medium. In the quantum limit, we have to read $q, x_i$ and the derived 
quantity $\xi$ as operators. 
We see both sides of open system dynamics --- Dissipation encoded in the damping kernel $\gamma$ and decoherence encoded in the noise term $\xi$. We can express $\gamma$ as
\begin{equation}
\gamma(t)=\sum_i \frac{c_i^2}{m_i\omega_i^2}\cos\omega_it
=\int_0^\infty \frac{d\omega}{\omega}J(\omega)\cos\omega t 
\label{eq:gammat}
\end{equation}
where we have introduced the spectral density of bath modes
\begin{equation}
J(\omega)=\sum_i \frac{c_i^2}{m_i\omega_i}\delta(\omega-\omega_i)
\end{equation}
which is the only quantifier necessary to describe the information encoded in the distribution of the $m_i$, $\omega_i$, and $c_i$. The right hand side of eq. \ref{eq:Langevin} is a noise term and reads
\begin{equation}
\xi(t)=\sum_i c_i\left[\left(x_i(0)-\frac{c_i}{m_i\omega_i^2}q(0)\right)\cos\omega_i t+\frac{p_i(0)}{m_i\omega_i}\sin\omega_i t\right].\label{eq:xit}
\end{equation}
This crucially depends on the initial condition of the bath. If we assume that the bath is initially 
equilibrated around the initial position $q(0)$ of the particle, we can show, using the standard
quantum-statistics of the simple harmonic oscillator, that the noise is unbiased, 
$\left\langle\xi(t) \right\rangle=0$, and its correlation function is given by
\begin{equation}
K(t)=\left\langle\xi(t)\xi(0)\right\rangle =\int d\omega J(\omega)\left[ \cos\omega t \left(2n(\hbar\omega)+1\right)-i\sin\omega t\right]
\label{eq:noisecorrelator}
\end{equation}
where $n$ is the Bose function, $n(\hbar\omega)=(e^{\hbar\omega/kT}-1)^{-1}$, and $2n(\hbar\omega)+1=\coth\left(\frac{\hbar\omega}{2k_B T}\right)$. Here and henceforth, angular brackets around an operator indicate the quantum-statistical 
average, $\langle O\rangle={\rm Tr} (\rho O)$ with $\rho$ being the appropriate density matrix. We will
get back to the topic of the initial condition in section \ref{nofactor} of this chapter. 

The noise described by $\xi(t)$ is the {\em quantum} noise of the bath. In particular, the correlation 
function is time-translation invariant, 
\begin{equation}
K(t)=\left\langle \xi(t+\tau)\xi(\tau)\right\rangle
\end{equation}
but not symmetric
\begin{equation}
K(-t)=\langle \xi(0)\xi(t) \rangle=K^\ast (t)\not=K(t).
\label{eq:correlator}
\end{equation}
which reflects the fact that $\xi$ as defined in eq. \ref{eq:xit} is a time-dependent operator which does generally 
not commute at two different times. Explicitly, the imaginary part of $K(t)$ changes its sign under time reversal. Indeed, if the derivation of eq. \ref{eq:noisecorrelator} is done explicitly, one directly sees that
it originates from the finite commutator. Moreover, we can observe that at $T\gg\omega$ we have
$2n+1\rightarrow 2k_BT/\hbar\omega\gg 1$, thus the integral 
in eq,\ \ref{eq:noisecorrelator} is dominated by the symmetric real part now describing purely thermal 
noise.  
At any temperature, the symmetrized semiclassical spectral noise power in frequency space reads
\begin{equation}
S(\omega)=\frac{1}{2}\left\langle \xi(t)\xi(0)+\xi(0)\xi(t) \right\rangle_\omega=S(-\omega)
\label{eq:somega}
%=J(\omega)\coth\left(\frac{\hbar\omega}{2k_BT}\right)
\end{equation}
where $\left\langle \dots\right\rangle_\omega$ means averaging and Fourier transforming. This
quantity contains a sign of the quantum nature of noise. Unlike classical noise, it does not disappear at low
temperatures $T\ll \hbar\omega/ k_B$, but saturates to a finite value set by the zero-point fluctuations,
whereas at high temperature we recover thermal noise. Note, that the same crossover temperature dictates the asymmetry in eq. (\ref{eq:correlator}). Both observations together can be identified with the
fact, that zero-point fluctuations only allow for emission of energy, not absorption, as will be 
detailed in a later section of this chapter. 

Our approach in this chapter is phenomenological. The main parameter of our model is the
spectral density $J(\omega)$. We will show in sections \ref{jcircuit1}, \ref{jcircuit2} and \ref{jcircuit3} how $J(\omega)$ can be derived explicitly
for Josephson junction circuits. Oscillator baths accurately model numerous other situations. Decoherence induced by phonons in quantum dot systems allows to directly identify the
phonons as the bath oscillators \cite{Brandes99,PRB052}, 
whereas in the case of electric noise from resistors or cotunneling in dots  \cite{PRBR041} it is less obvious --- the Bosons are electron-hole excitations, which turn out to 
have the commutation relation of hard-core bosons \cite{Vondelft98} with the hard-core term being of little effect in the limits of interest \cite{Weiss99}. 

Going back to our
phenomenology, we introduce the most important case of an Ohmic bath
\begin{equation}
J(\omega)=\gamma\omega f(\omega/\omega_c).
\label{eq:ohmic}
\end{equation}
Here, $\gamma$ is a constant of dimension frequency and $f$ is a high-frequency cutoff function providing $f(x)\simeq1$ at $x<1$ and $f\rightarrow 0$ at $x>1$. Popular choices
include the hard cutoff, $f(x)=\theta(1-x)$, exponential cutoff, $f(x)=e^{-x}$, and the Drude cutoff
$f(x)=\frac{1}{1+x^2}$. We will see in section \ref{jcircuit1} that the Drude cutoff plays a significant role in finite electrical circuits, 
so we chose it here for illustration purposes. In this case, the damping kernel reduces to 
\begin{equation}
\gamma(\tau)=\gamma\omega_c e^{- \omega_c\tau}.
\end{equation}
 For 
$\omega_c\rightarrow\infty$, $\gamma$ becomes a delta function and we recover the classical damping with damping constant $\gamma$, $\gamma(\tau)=\gamma\delta(\tau)$. Here, 	``classical damping" alludes to the damping of particle
motion in  fluid or of charge transport in a resistor (thus the name Ohmic, see also section \ref{jcircuit1}). 
With finite $\omega_c$, 
the Ohmic models leads to classical, linear friction proportional to the velocity, smeared out over
a memory time set by the inverse cutoff frequency defining a correlation time $t_c=\omega_c^{-1}$. On the other hand, as it turns out in the analysis of the model e.g. in section \ref{sbredfield}, an infinite cutoff always leads to unphysical divergencies. Examples will be given later on. All examples from the class of superconducting qubits have a natural ultraviolet cutoff set by an appropriate
$1/RC$ or $R/L$ with $R$, $L$, and $C$ being characteristic resistances, inductances, and capacitances of the circuit, respectively. Note, that parts of the open quantum systems literature do not make this last observation.

We will not dwell on methods of solution\ of the quantum Langevin equation, as the focus of this work
is the decoherence of qubit systems. Methods include the associated Fokker-Planck equation, path integrals, and quantum trajectory simulations. The quantum Langevin equation finds application in the theory of quantum decay in chemical reactions, the dissipative harmonic oscillator, and the decoherence of double-slit experiments. 

\subsubsection{How general are oscillator baths?
\label{linearresponse}}
Even though the model introduced looks quite artificial and specific, it applies to a broad range of systems. The model essentially applies as long as the heat
bath can be treated within linear response theory, meaning that it is essentially infinite (i.e. cannot be 
exhausted), has a regular spectrum, and is in thermal equilibrium. We outline the requirement
of only weakly perturbing the system, i.e. of linear response theory \cite{Kubo91}. The derivation is rather sketchy
and just states the main results because this methodology will not be directly used later on. 
Introductions can be found e.g. in Ref. \cite{Kubo91,Ingold98,Callen51}

In linear response theory, we start from a Hamiltonian $H_0$ of the oscillator bath which is perturbed by
an external force $F$ coupling to a bath operator $Q$,
\begin{equation}
H=H_0-FQ
\end{equation}
where the perturbation must be weak enough to be treated to lowest order. It is a result of linear response
theory that the system responds by a shift of $Q$ (taken in the Heisenberg picture) according to
\begin{equation}
\left\langle\delta Q(t)\right\rangle_\omega=\chi(\omega)F(\omega)
\end{equation}
where the susceptibility $\chi$ can be computed to lowest order as the correlation function
\begin{equation}
\chi=\left\langle Q(0)Q(0)\theta(t)\right\rangle_\omega
\end{equation}
computed in thermal equilibrium. We can split the correlation function into real and imaginary parts
$\chi=\chi^\prime+i\chi^{\prime\prime}$. The real part determines the fluctuations, i.e.
\begin{equation}
\frac{1}{2}\left\langle \delta Q(t)\delta Q(0)+\delta Q(0) \delta Q(t)\right\rangle_\omega = \chi^\prime
\end{equation}
whereas the imaginary part determines the energy dissipation
\begin{equation}
\left\langle E(t) \right\rangle_\omega=\omega\chi^{\prime\prime}|F|^2.
\end{equation}
together with the equations \ref{eq:gammat} and \ref{eq:somega} tracing both damping and noise back to  a single function $\chi$ constitute 
the famous fluctuation-dissipation theorem \cite{Callen51}, a generalization of the Einstein relation in diffusion.

In this very successful approach we have characterized the distribution of the observable $Q$ close
to thermal equilibrium by its two-point correlation function alone. This is a manifestation of the fact
that its distribution, following the central limit theorem is Gaussian, i.e. can be characterized by two
numbers only: mean and standard deviation. Oscillator baths provide exactly these ingredients: by properly
chosing $J(\omega)$ they can be fully adjusted to any $\chi(\omega)$, and all higher correlation functions --- correlation functions involving more than two operators --- can also be expressed through $J$ hence do not contain any independent piece of information. 

This underpins the initial statement that oscillator baths can describe a broad range of environments, 
including those composed of Fermions and not Bosons, such as a resistor. As explained in section \ref{qbm}, the oscillators are introduced artifically --- on purely statistical grounds as a tool
to describe fluctuations and response --- and can only sometimes be directly identified with a physical entity.

There are still a number of environments where the mapping on an oscillator bath is in general not
correct. These include  i) baths of uncoupled spins (e.g. nuclear spins),
 which are not too big and can easily saturate, i.e. explore the full finite capacity of their bounded energy spectrum ii) shot noise, which is not in thermal equilibrium iii) nonlinear electrical circuits such as many Josephson circuits and iv) in most cases $1/f$ noise, whose microscopic explanation either
hints at non-Gaussian (spin-like) or nonequilibrium sources as discussed in section \ref{onef}. 

\subsubsection{Oscillator bath models for Josephson junction devices \label{jcircuit1}}

We have now learned two approaches to characterize the oscillator bath: through noise, and through friction. We will now apply the characterization by friction to a simple Josephson circuit with Josephson energy $E_{\rm J}$, junction capacitance $C_{\rm J}$ and arbitrary shunt admittance in parallel, all biased by an external current $I_{\rm B}$. We are extending
the method presented in the previous chapter \cite{Geller06} to include the admittance. We start with 
the elementary case of a constant conductance, $Y(\omega)=G$. The total current splits up into the three elements as
\begin{equation}
I_B=I_c \sin\phi+C\frac{\Phi_0}{2\pi}\ddot{\phi}+G\frac{\Phi_0}{2\pi}\dot{\phi}.
\end{equation}
Reordering terms, we can cast this into the shape of Newton's equation for a particle with coordinate $\phi$. 
\begin{equation}
C\left(\frac{\Phi_0}{2\pi}\right)^2\ddot{\phi}+G\left(\frac{\Phi_0}{2\pi}\right)^2\dot{\phi}+\frac{\partial V}{\partial\phi}=0.
\label{eq:currentconservation}
\end{equation}
Here, we have multiplied the equation by another $\Phi_0/2\pi$ to ensure proper dimensions of the 
potential energy
\begin{equation}
V(\phi)=-I_B\frac{\Phi_0}{2\pi}\phi+E_J(1-\cos\phi)
\end{equation}
where we have introduced the Josephson energy $E_J=I_0\Phi_0/2\pi$. 
This expression can be readily compared to eq. \ref{eq:Langevin}. We see that the friction term has no 
memory, i.e. $\gamma(t)\propto \delta(t)$,  and using the results of section \ref{qbm} we can infer that $J(\omega)=G(\Phi_0/2\pi)\omega$, i.e.\ an Ohmic resistor leads naturally to an Ohmic
spectral density as mentioned before. Note that this has no cutoff, but any  model of an Ohmic resistor leads to reactive behavior at high frequencies.

We see that we missed the noise
term on the right, which would represent current noise originating in $G$ and which would have to be included in a more sophisticated circuit analysis which careful engineers would do. By applying the fluctuation dissipation theorem to $\gamma$ we can add on the proper noise term, whose correlation function 
is given by equation (\ref{eq:noisecorrelator}) --- or we can simply use this equation with the 
$J(\omega)$ obtained. 

We want to generalize this system now to an arbitrary shunt admittance $Y(\omega)$. For that, 
it comes in handy to work in Fourier space and we denote the Fourier transform by $\mathcal{F}$. Analogous to eq.  (\ref{eq:currentconservation}), we can find the following expression
\begin{equation}
-\omega^2C\left(\frac{\Phi_0}{2\pi}\right)^2\phi+i\omega Y(\omega)\left(\frac{\Phi_0}{2\pi}\right)^2\phi+\mathcal{F}\left(\frac{\partial V}{\partial\phi}\right)=0.
\label{eq:newton}
\end{equation}
We have to remember that the damping Kernel $\gamma$ is the Fourier cosine transform of $J(\omega)/\omega$, which also implies that it is a real valued function. We can split $Y$ into real (dissipative) and imaginary (reactive) parts $Y=Y_d+iY_r$. For any finite electrical circuit, $Y_d$ is always an even and $Y_r$ always an odd function of frequency. All this allows us to rewrite eq.\ \ref{eq:newton}
\begin{equation}
-\omega^2C\left(\frac{\Phi_0}{2\pi}\right)^2\phi-\omega Y_r(\omega)\left(\frac{\Phi_0}{2\pi}\right)^2\phi+i\omega Y_d(\omega)\left(\frac{\Phi_0}{2\pi}\right)^2\phi+\mathcal{F}\left(\frac{\partial V}{\partial\phi}\right)=0.
\label{eq:colouredbrownian}
\end{equation}
Thus, the general expression for the spectral density reads $J(\omega)=\omega Y_d=\omega {\rm Re}Y(\omega)$, i.e. it is controlled by the {\em dissipative} component of $Y(\omega)$ alone. There is a new term containing the reactive component $Y_r$ which modifies the non-dissipative part of the dynamics and can lead e.g. to mass or potential renormalization, or something more complicated. Comparing this result to the structure of the susceptibility $\chi$ in the discussion of section \ref{linearresponse} it looks like the real and imaginary part have changed their role and there is an extra factor of $\omega$. This is
due to the fact that $Y$ links $I$ and $V$, whereas the energy-valued perturbation term in the sense
of section \ref{linearresponse} is $QV$. This aspects adds a time-derivative $Y=\dot{\chi}$ which leads
to a factor $i\omega$ in Fourier space. 

This last result can be illustrated by a few examples. If $Y(\omega)=G\Phi_0/2\pi$, we recover the previous equation (\ref{eq:currentconservation}). If the shunt is a capacitor $C_s$, we have $Y(\omega)=i\omega C_s$ and we get from eq. (\ref{eq:colouredbrownian}) the equation of motion of a particle with 
larger mass, parameterized by a total capacitance $C_{\rm tot}=C_{\rm J}+C_{\rm s}$. On the other
hand, if the shunt is an inductance $L_{\rm s}$, we obtain $Y(\omega)=(i\omega L_{\rm s})^{-1}$, leading to a new contribution to the potential originating from the inductive energy
\begin{equation}
V_{tot}(\phi)=V(\phi)+\frac{(\Phi_0)^2}{8\pi^2 L}\phi^2
\end{equation}
and no damping term. Finally, let us consider the elementary mixed case of a shunt consisting of a resistor $R_s$ and a capacitor $C_s$ in series. We find $Y(\omega)=\frac{i\omega C_s}{1+i\omega R_sC_s}$ which can be broken into a damping part which is supressed below a rolloff frequency $\omega_r=(RC)^{-1}$, $Y_d\frac{1}{R}\frac{1}{1+\omega^2/\omega_r^2}$ and a reactive part 
which responds capacitively below that rolloff, $Y_r=i\omega C\frac{1}{1+\omega^2/\omega_r^2}$. As the rolloffs are very soft, there is no straightforward mapping onto a very simple model and we have to accept that the dynamics get more complicated and contain a frequency-dependent mass and friction as well as
time-correlated noise, all of which gives rise to rich physics \cite{PRB051}.

\section{Single qubit decoherence}

\subsection{Two-state oscillator bath models}

In the previous section, we introduced the notion of an oscillator bath environment for continuous systems including biased Josephson junctions. We derived quantum Langevin equation demonstrating the analogy to classical dissipative motion, but did not describe how to solve them. In fact, solving these equations in all generality is extremely hard in the quantum limit, thus a restriction of generality
is sought. For our two-state systems (TSS) of interest, qubits, we are specifically interested in the case where the potential in the Hamiltonian of eq.\ \ref{eq:hamiltonian_qbm} forms a double well with exactly one bound state
per minimum, tunnel-coupled to each other and well separated from the higher excited levels,
\cite{Geller06}. When we also concentrate on the low-energy dynamics, we can replace the particle coordinate $q$ by $q_0\sigma_z$ and the Hamiltonian reads
\begin{equation}
H=\frac{\epsilon}{2}\sigma_z+\frac{\Delta}{2}\sigma_x+\frac{\sigma_z}{2}\sum_i \lambda_i(a_i+a_i^\dagger)+\sum_i \omega_i(a_i^\dagger a_i +1/2),
\label{eq:hamiltonian_tss}
\end{equation}
where $\epsilon$ is the energy bias and $\Delta$ is the tunnel splitting. This is the famous Spin-Boson Hamiltonian \cite{Leggett87,Weiss99}. We have dropped the counterterm, which is $\propto q^2$ in the continuous limit and, due to $q=\pm q_0$ is constant in the two-state case. The spectral density is 
constructed out of the $J(\omega)$ in the continuous limit
\begin{equation}
J_{\rm TSS}=\sum_i\lambda_i^2\delta(\omega-\omega_i)=\frac{q_0^2}{2\pi\hbar}J(\omega)
\label{eq:hamiltonian:TSS}
\end{equation}
The Spin-Boson Hamiltonian, eq.\ (\ref{eq:hamiltonian:TSS}) is more general than the truncation of the energy spectrum in a double-well potential may suggest. In fact, it can be derived by an alternative procedure which performs the two-state approximation first (or departs from a two-state Hamiltonian without asking for its origin) and then characterizes the bath. The oscillator bath approximation holds under the same conditions explained in 
section \ref{linearresponse} The Spin-Boson model makes the assumption, that each oscillator 
couples to the same observable of the TSS which can always be labelled $\sigma_z$. This is a 
restrictive assumption which is not necessarily true for all realizations of a dissipative two-state system. 

As the two-state counterpart to classical friction used in the continuous case is not straightforward to determine, the environmental spectrum is computed from the semiclassical noise of the environment, following the prescription that, if we rewrite 
eq.\ \ref{eq:hamiltonian_tss} in the interaction picture with respect to the bath as
\begin{equation}
H_I=\frac{\epsilon+\delta\epsilon(t)}{2}\sigma_z+\frac{\Delta}{2}\sigma_x +
\sum_i \omega_i(a_i^\dagger a_i +1/2)
\label{eq:hamiltonian_tss_noise}
\end{equation}
we can identify $\delta\epsilon$ for any physical model mapping on the Spin-Boson model as
\begin{equation}
\frac{1}{2}\left\langle\delta\epsilon(t)\delta\epsilon(0)+\delta\epsilon(0)\delta\epsilon(t)\right\rangle_\omega=J_{\rm TSS}(\omega)\coth\left(\frac{\hbar\omega}{2k_B T}\right)
\end{equation}
An application of this procedure will be presented in the next subsection.

\subsubsection{Characterization of qubit environments through noise \label{jcircuit2}}

\begin{figure}[htb]
\includegraphics[width=0.9\columnwidth]{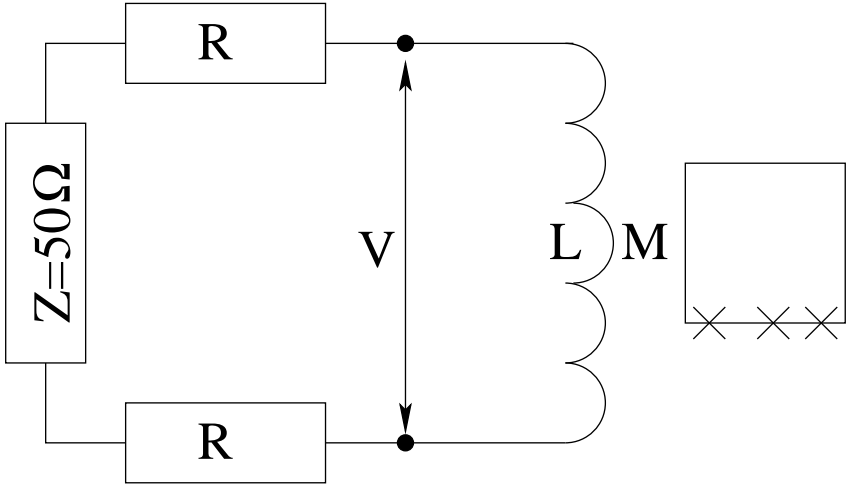}
\caption{Typical on-chip electromagnetic environment of a superconducting flux qubit, consisting
of the flux control coil with self-inductance $L$, mutual inductance $M$ to the qubit, shunt impedance $Z$ and on-chip decoupling resistors $R$.\label{fig:biasingcircuit}}
\end{figure}

A standard application of the characterization of the environment is the description of control electronics of relatively modest complexity, attached to a flux qubit. We look at the definite example shown in
 \ref{fig:biasingcircuit}. It shows a simplified model of the microwave leads providing the control of 
a flux qubits. The microwaves inductively couple to the sample by a mutual inductance $M$ between the qubit and a coil with self-inductance $L$. These leads are mounted in the cold part of the cryostat, usually on the qubit chip, and are connected to the outside world by a coaxial line which almost inevitably has 
an impedance of $Z=50\Omega$. That impedance provides --- in light of the discussion in the previous section --- a significant source of damping and decoherence. As a design element, one can put two resistors of size $R$ close to the coil. 

The environmental noise is easily described by the Nyquist noise \cite{Callen51} of the voltage $V$ between the arms of the circuit, see figure \ref{fig:biasingcircuit}. The Johnson-Nyquist formula gives the voltage noise
\begin{equation}
S_V=\frac{1}{2}\left\langle V(t)V(0)+V(0)V(t)\right\rangle=\hbar\omega {\rm Re} Z_{\rm eff}
\coth\left(\frac{\hbar\omega}{2k_BT}\right)
\label{eq:nyquist}
\end{equation}
where $Z_{\rm eff}$ is the effective impedance between the arms, here of a parallel setup of
a resistor and an inductor
\begin{equation}
Z_{\rm eff}=\frac{i\omega L_{\rm eff} R}{R+i\omega L_{\rm eff}},
\end{equation}
and $L_{\rm eff}$ is the total impedance of the coupled set of conductors as seen from the circuit. For 
microwave leads, the total inductance is dominated by the self-inductance of the coil, hence $L_{\rm eff}\approx L$.

We need to convert the voltage noise into energy level noise of the qubit. A voltage fluctuation $\delta V$ leads to a current fluctuation in the coil following
\begin{equation}
\delta I=\delta V/i\omega L.
\end{equation}
The current noise produces flux noise through the qubit loop
\begin{equation}
\delta\Phi=M\delta I=\frac{M}{i\omega L}\delta V
\end{equation}
which converts into energy bias noise following
\begin{equation}
\delta\epsilon=I_s\delta\Phi=\frac{MI_s}{i\omega L}\delta V
\end{equation}
with $I_s$ being the circulating current in the potential minima of the qubit. Thus, the energy level correlation function reads
\begin{equation}
S_\epsilon=\left(\frac{MI_s}{i\omega L}\right)^2S_V
\end{equation}
which allows us to express the spectral density through the impedance as
\begin{equation}
J(\omega)=\hbar\omega \left(\frac{MI_s}{i\omega L}\right)^2 {\rm Re} Z_{\rm eff}(\omega).
\label{eq:jthroughz}
\end{equation}
With the specific circuit shown in figure \ref{fig:biasingcircuit}, we find that the environment is Ohmic
with a Drude cutoff
\begin{equation}
J(\omega)=\frac{\alpha\omega}{1+\omega^2/\omega_c^2}
\end{equation}
with $\omega_c=L/R$ and $\alpha=\frac{4M^2I_s^2}{h(Z+2R)}$. Thus, we find a simple method to engineer the decoherence properties of thw circuit with our goal being to reduce $J(\omega)$ by decoupling the device from the shunt $Z$. The method of choice is to put large resistors $R$ on chip. Their size will ultimately be limited by the necessity of cooling them to cryogenic temperatures. The friction method introduced earlier, section \ref{jcircuit1} , leads to the same result.

\subsubsection{Linearization of nonlinear environments \label{jcircuit3}}

In general, nonlinear environments important for qubit devices can also be identified. In superconducting devices, these include electronic environments which in addition to the linear circuit elements discussed in the previous section, also contain Josephson junctions. In general, such environments cannot be described by oscillator bath models, whose response would be strictly linear. Here, we want to concentrate on the case of a nonlinear environment --- a SQUID detector --- in the regime of small signal response, i.e. in a regime where it can be linearized. This linearization can be illustrated by the concept of Josephson inductance. Let us remind ourselves, that a linear inductor is defined through the following current-flux relation
\begin{equation}
I(\Phi)=\Phi/L
\end{equation}
whereas the small flux-signal response of a Josephson junction can be approximated as
\begin{equation}
I=\sin\left( 2\pi\frac{\Phi}{\Phi_0}\right)\simeq I_c\sin\left(2\pi\frac{\bar{\Phi}}{\Phi_0}\right)+\frac{\delta\Phi}{L_J}
\end{equation}
where we have split the flux into its average $\bar{\Phi}$ and small deviations $\delta\Phi$ and
have introduced the Josephson inductance $L_{\rm J}=\Phi_0/2\pi I_c\cos\bar{\phi}$. 
Thus, the small-signal response is inductive. 

\begin{figure}[htb]
\includegraphics[width=0.9\columnwidth]{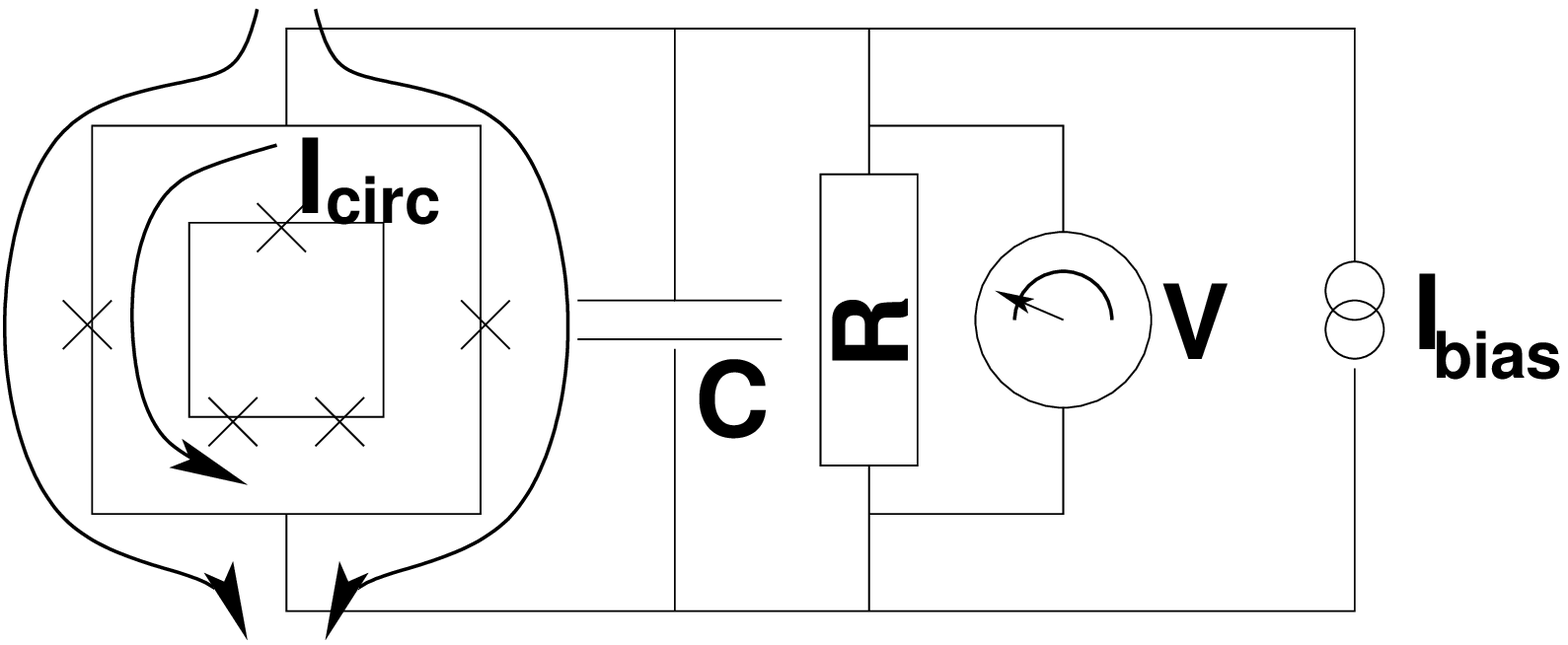}
\includegraphics[width=0.9\columnwidth]{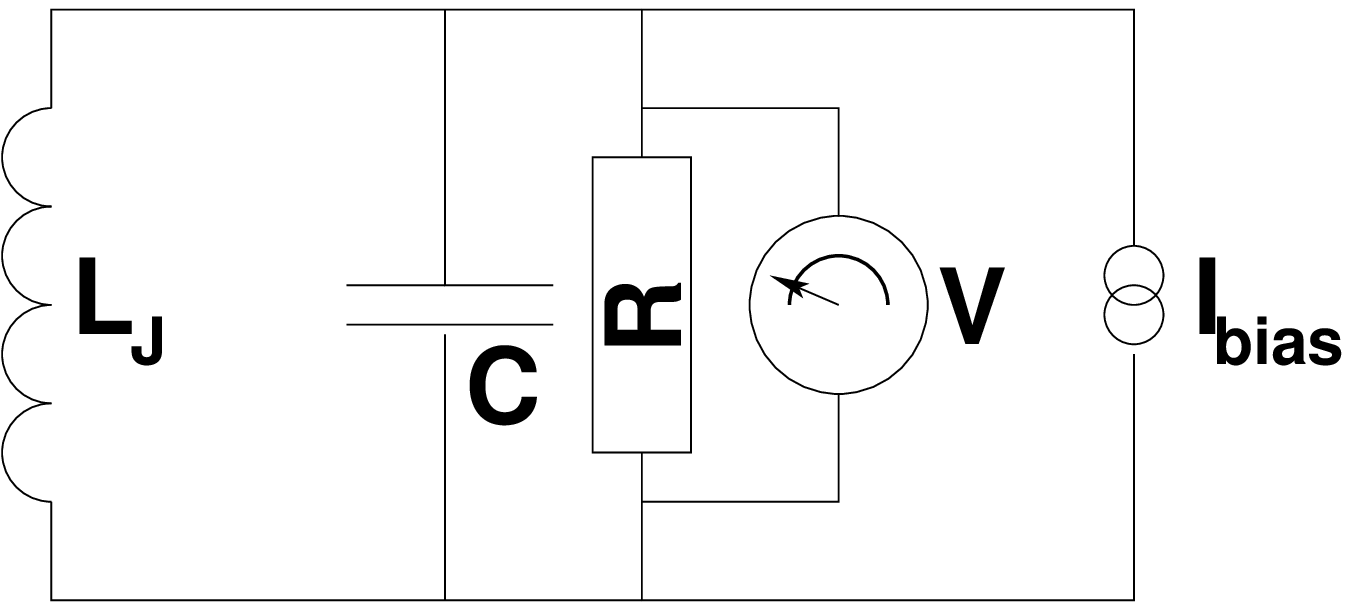}
\caption{Upper panel: DC-SQUID readout circuit consisting of  the actual SQUID, a shunt capacitor, and a voltmeter with an unavoidable resistor. Lower panel: Linearized circuit used for the noise calculation \label{fig:squid}}
\end{figure}

We would now like to demostrate this idea on the example of a DC-SQUID detector inductively coupled to the qubit, see fig. \ref{fig:squid}.

In the first stage, we again need to find the voltage noise between the branches of the circuit. This is given by eq.\ (\ref{eq:nyquist}) with the appropriate inductance calculated from the cicruit shown in the lower panel of fig.\ \ref{fig:squid}, $Z_{\rm eff}^{-1}=R^{-1}+i\omega C+(i\omega L_J)^{-1}$. This is the impedance of an $LC$ resonator with damping. The conversion into energy level noise goes along similar lines as before, incorporating the SQUID equations as described here and in standard
literature \cite{Tinkham96,Clarke04}. 

The DC-SQUID is a parallel setup of two Josephson junctions $1$ and $2$, which for simplicity are assumed to be identical. The total current flowing through the device is 
\begin{equation}
I_B=I_c(\sin\phi_1+\sin\phi_2)=2I_c\cos (\delta\phi/2)\sin\bar{\phi}
\label{eq:biascurrent2}
\end{equation}
where we have introduced $\bar{\phi}=(\phi_1+\phi_2)/2$ and $\delta\phi=\phi_1-\phi_2$. Now we need to remember that the phases $\phi_i$ are connected to the Schr\"odinger equation for the superconducting condensate. Thus, an elementary calculation \cite{Tinkham96,Clarke04} leads to
\begin{equation}
\delta\phi=2\pi\frac{\Phi}{\Phi_0} {\rm mod} 2\pi
\end{equation}
where $\Phi$ is the total magnetic flux through the loop. This is identical to the flux applied externally using a biasing coil plus the qubit flux as we neglect self-inductance. Thus, for the bias
current $I_B$ the DC-SQUID acts like a tunable Josephson junction with a critical current $I_{\rm c, eff}(\Phi)=2I_c|\cos(\pi\Phi/\Phi_0)|$. Thus, we can translate voltage fluctuations into phase fluctuations as
\begin{equation}
\delta\bar{\phi}=\left(\frac{2\pi}{\Phi_0}\right)\frac{\delta V}{i\omega}.
\end{equation}
The qubit is coupling to the magnetic flux which --- assuming a symmetric SQUID geometry - is coupled only to the circulating current
\begin{equation}
I_{\rm circ}=I_c(\sin\phi_1-\sin\phi_2)/2=I_c\cos (\bar\phi)\sin\frac{\pi\Phi}{\Phi_0}.
\label{eq:biascurrent}
\end{equation}
We can now express its fluctuations through the fluctuations of $\bar{\phi}$
\begin{equation}
\delta I_{\rm circ}=-I_c\sin\frac{\pi\Phi}{\Phi_0}\sin(\bar\phi)\delta\bar\phi=\frac{I_B}{2}\tan\frac{\pi\Phi}{\Phi_0}\delta\phi
\label{eq:circcurrent}
\end{equation}
where in the last step we  have used eq. \ref{eq:biascurrent2}. With the remaining steps analogous to the
previous section, we obtain
\begin{equation}
J(\omega)=\hbar\omega \left(M I_s\frac{I_B}{2}\frac{2\pi}{\Phi_0}\tan\frac{\pi\Phi}{\Phi_0}\right)^2 {\rm Re} Z_{\rm eff}.
\end{equation}
Here, $Z_{\rm eff}$ is the impedance of the linearized circuit shown in the bottom panel of fig.\ \ref{fig:squid}.
This result reveals a few remarkable features. Most prominently, it shows that $J(\omega)$ can be tuned by shifting the working point of the linearization through changing the bias current $I_B$. In particular, $J(\omega)$ can be set to zero by chosing $I_B=0$. The origin of this decoupling can be seen in 
eq.\ \ref{eq:circcurrent}, which connects the bias current noise to the circulating current noise. The physical reason for this is, that in the absence of a bias current the setup is fully symmetric --- any noise
from the external circuitry splits into equal amounts on the branches of the loop and thus does not lead to flux noise. For a detector, this is a highly desired property. It allows to switch the detector off completely. When we do not bias, we have (for the traditional switching current measurement) no senitivity and with
it no backaction. This means, that if the device is really highly symmetric, one can push this device to the 
strong measurement regime while still being able to operate in the "off" state of the detector.
This effect has been predicted in Refs. \cite{EPJB03,ASSP03}. Experimentally, it was first observed
that the decoupled point was far from zero bias due to a fabrication issue \cite{Burkard05}, which was later
solved such that our prediction has indeed been verified \cite{Bertet05}. 

\subsubsection{The Bloch equation\label{ch:bloch}}

So far, we have discussed the characterization of the environment at length. We did not specify how
to describe the qubit dynamics under its influence. For a continuous system, we have derived the quantum Langevin equation
(\ref{eq:Langevin}). Even though this eqution looks straightforward, solving it for potentials others than the harmonic oscillator is difficult without further approximations. We will now show first how to describe decoherence in a phenomenological way and then discuss how to reconcile microscopic modelling with the Bloch equation. 

For describing the decoherence of a qubit we have to use the density matrix formalism. which can describe pure as well as mixed states. In the case of a qubit with a two-dimensional Hilbert space, we can fully parameterize the density matrix by its three spin projections
$S_i={\rm Tr} (\rho\sigma_i)$, $i=x,y,z$ as
\begin{equation} 
\rho=\frac{1}{2}\left(1+\sum_i S_i\sigma_i\right)
\end{equation}
where the $\sigma_i$ are Pauli matrices. This notation is inspired by spin resonance and is applicable
to any two-state system including those realized in superconducting qubits. We can take the analogy 
further and use the typical NMR notation with a strong static magnetic field $B_z(t)$ applied in one direction identified as the $z$-direction and a small AC field, $B_x(t)$ and $B_y(t)$ in the $xy$-plane. In that case, there is clearly a preferred-axis symmetry and two distinct relaxation rates, the longitudinal rate $1/T_1$ and the transversal rate $1/T_2$ can be introduced phenomenologically to yield
\begin{eqnarray}
\dot{S}_z&=&\gamma(\vec{B}\times\vec{S})_z-\frac{S_z-S_{z,eq}}{T_1}\\
\dot{S}_{x/y}&=&\gamma(\vec{B}\times\vec{S})_{x/y}-\frac{S_{x/y}}{T_2}
\end{eqnarray}
where we have introduced the equilibrium spin projection $S_{z,eq}$ and the spin vector $\vec{S}=(S_x, S_y, S_z)^{T}$. Note that the coherent part of the time evolution is still present. It enters the Bloch 
equation via the Hamiltonian, decomposed into Pauli matrices as $H= -\gamma\vec{B}\cdot\vec{S}$.  This spin notation is also useful for superconducting qubits, even though the three components usually depend very distinct observables such as charge, flux, and current. This parameterization leads to the practical visualization of the state and the Hamiltonian as a point and an axis in three-dimensional space respectively. The free evolution of the qubit then corresponds  to Larmor precession around the magnetic field.  The pure states of the spin have $\vec{S}^2=1$ and are hence on a unit sphere, the Bloch sphere, whereas the mixed states are inside the sphere --- in the Bloch ball. 

The rates are also readily interpreted in physical terms. As the large static 
field points in the $z$-direction in our setting, the energy dissipation is given as
\begin{equation}
\frac{d\langle E\rangle}{dt}=-\gamma B_z\dot{S_z}
\end{equation}
and hence its irreversible part is given through $1/T_1$. On the other hand, the purity (or linearized entropy) $P={\rm Tr}\rho^2 =1/4+\sum_i S_i^{2}$ decays as
\begin{equation}
\dot{P}=2\sum_i \dot{S}_i S_i=-\frac{S_x^2+S_y^2}{T_2}-\frac{S_z(S_z-S_{z,eq})}{T_1}
\end{equation}
thus all rates contribute to decoherence. Note, that at low temperatures $S_{z,eq}\rightarrow 1$ so the $T_1$-term in general augments the purity and reestablishes coherence. This can be understood as the system approaches the ground state, which is a pure state. In this light, it needs to be 
imposed that 
$P\le1$ as otherwise the density matrix has negative eigenvalues. This enforces $T_2\le 2T_1$. 

\subsection{Solutions of the Bloch equation and spectroscopy}

The rates shown in the Bloch equation are also related to typical spectrocopic parameters 
\cite{Abragam83,PRB03}. We chose 
a rotating driving field
\begin{eqnarray}
B_x&=&(\omega_R/\gamma)\cos\omega t\\
B_y&=&(\omega_R/\gamma)\sin\omega t.
\end{eqnarray}
In spectroscopy, we are asking for the steady state population, i.e.\ for the long-time limit of 
$S_z$. Transforming the Bloch equation into the frame co-rotating with the driving field and computing
the steady-state solution, we obtain
\begin{equation}
S_z(\omega)=\frac{\omega_R^2}{(\omega-\gamma B)^2+\gamma^2}
\end{equation}
with a linewidth $\gamma^2=1/T_2^2+\omega_R^2 T_2/T_1$. This simple result allows spectroscopic 
determination of all the parameters of the Bloch equation: At weak driving, $\omega_R\sqrt{T_1T_2}\ll1$, the line width is $1/T_2$. This regime can be easily identified as the spectral line not being saturated, i.e. the height grows with increasing drive.  In fact, the height of the resonance is $S_z(\gamma B)=\omega_R^2T_2^2$, which (knowing  $1/T_2$) allows to determine $\omega_R$. Due to the
heavy filtering between the room-temperature driving and the cryogenic environment, 
this is not known a priory.
To determine $T_1$, one goes to the
high driving regime with a saturated line, i.e. a line which does not grow any more with higher power, 
$\omega_R\sqrt{T_1T_2}\gg 1$ and finds a line width of
$\omega_R\sqrt{T_1/T_2}$. With all other parameters known already, this allows to find $T_1$. Using this approach is helpful to debug an experiment which does not work yet. Alternatively, real-time measurements of $T_1$ are possible under a wide range of conditions. 

\subsubsection{How to derive the Bloch equation: The Bloch-Redfield technique \label{ch:redfield}}

We now show how to derive Bloch-like equations from the system-bath models we studied before using 
a sequence of approximations. The {\em Born approximation} works if the coupling between
system and bath is weak.  The {\em Markov approximation} works if the coupling between system and
bath is the slowest process in the system, in particular if it happens on a time scale longer than the
correlation time of the environment.  Quantitatively, we can put this into
the {\em motional narrowing condition}
\begin{equation}
\frac{\lambda\tau_c}{\hbar} \ll 1 \ , 
\label{eq:narrow}
\end{equation} 
where $\lambda$ is the coupling strength between the system and its environment and
$\tau_c$ the correlation time of the environment. In the case treated in eq. \ref{eq:ohmic} we would have
$\tau_c=1/\omega_c$.
If this is satisfied, an averaging
process over a time scale longer than $\tau_c$ but shorter than $\lambda^{-1}$ can lead to simple evolution
equations, the so-called Bloch-Redfield equations \cite{Argyres64}. The derivation
in Ref.~\cite{Cohen92} follows this inspiration. 
We will follow the very elegant and rigorous derivation using projection operators as 
given in \cite{Argyres64,Weiss99}. We are going to look at a quantum subsystem with an arbitrary finite
dimensional Hilbert space, accomodating also qudit and multiple-qubit systems. 

As a starting point for the derivation of the Bloch-Redfield equations
(\ref{redfield}), one usually \cite{Weiss99} takes the Liouville equation
of motion for the density matrix of the whole system $W(t)$ (describing the 
time evolution of the system)
\begin{equation}
\dot{W}(t)=-\frac{i}{\hbar} \left[ H_{\rm total},W(t) \right] = \mathcal{L}_{\rm total}W(t)\ , \label{Liouville}
\end{equation} 
where 
$H_{\rm total}$ is the total Hamiltonian and $\mathcal{L}_{\rm total}$ the total Liouvillian of the whole system. This  notation of the Liouvillian uses the concept of a {\em superoperator}. Superoperator
space treats density matrices as vectors. Simply arrange the matrix elements in a column, and each 
linear operation on the density matrix can be written as a (super)matrix multiplication. Thus, the right hand side of the Liouville 
equation can be written as a single matrix products, not a commutator, where a matrix acts from
the left and the right at the same time.  Hamiltonian and Liouvillian consist
of parts for the relevant subsystem, the reservoir and the interaction between these
\begin{eqnarray}
H_{\rm total} & = & H_{\rm sys}+H_{\rm res}+H_I \\
\mathcal{L}_{\rm total} & = & \mathcal{L}_{\rm sys} + \mathcal{L}_{\rm res} + \mathcal{L}_I.
\label{eq:LiouvilleSplit}
\end{eqnarray}

$H_{\rm sys}$ is the Hamiltonian which describes the quantum system (in our case: 
the qubit setup), $H_{\rm res}$ represents for the environment
and $H_I$ is the interaction Hamiltonian between system and bath. \\
Projecting the density matrix of the whole system $W(t)$ on the relevant
part of the system (in our case the qubit), one finally gets
the reduced density matrix~$\rho$ acting on the quantum system alone
\begin{equation}
\rho(t)={\rm Tr}_B W(t)=PW(t) \ , \label{rho}
\end{equation}
so $P$ projects out onto the quantum subsystem.
As in the previous derivation in section \ref{qbm}, we need to formally solve the irrelevant part of the Liouville equation 
first. Applying $(1-P)$, the projector on the irrelevant part, to eq. \ref{Liouville} and the obvious
$W=PW+(1-P)W$
we get
\begin{equation}
(1-P)\dot{W}=(1-P)\mathcal{L}_{\rm total}(1-P)W+(1-P)\mathcal{L}_{\rm total}\rho.
\end{equation}
This is an inhomogenous linear equation of motion which can be solved with variation of constants, 
yealding
\begin{equation}
(1-P)\rho(t) =\int_0^t\limits dt'  e^{(1-P)\mathcal{L}_{\rm total}(t-t)'} (1 - P) \mathcal{L}_{\rm total} \rho(t') + e^{(1-P)\mathcal{L}_{\rm total}t}(1 - P) W(0). 
\end{equation}
Putting this result into equation (\ref{Liouville}) one gets the 
Nakajima-Zwanzig equation \cite{Nakajima58,Zwanzig60}
\begin{eqnarray}
\dot{\rho}(t) & = &P \mathcal{L}_{\rm total}\rho(t) + \int_0^t\limits dt' P \mathcal{L}_{\rm total} e^{(1-P)\mathcal{L}_{\rm total}(t-t')} (1 - P) \mathcal{L}_{\rm total} \rho(t') + \nonumber \\
& & + P \mathcal{L}_{\rm total} e^{(1-P)\mathcal{L}_{\rm total}t}(1 - P) W(0). 
\end{eqnarray}
So far, all we did was fully exact. 
The dependence on the initial value of the irrelevant part of the density operator $(1-P)W(0)$ is
dropped, if the projection operator is chosen appropriately -- using factorizing initial conditions, i.e. $W=\rho\otimes (1-P)W$. A critical assesment of this assumption will be given in section \ref{nofactor}.
As $P$ commutes with
$\mathcal{L}_{\rm sys}$, one finds
\begin{equation}
\dot{\rho} = P(\mathcal{L}_{\rm sys} + \mathcal{L}_I)\rho(t) + \int_0^t\limits dt' P \mathcal{L}_I e^{(1-P)\mathcal{L}_{\rm total}(t-t)'} (1-P)\mathcal{L}_I \rho(t') \label{eq:nz2}.
\end{equation}
The reversible motion of the relevant system is described by the first (instantaneous)
term of eq.~(\ref{eq:nz2}), which contains the system Hamiltonian in $\mathcal{L}_{\rm sys}$ and 
a possible global energy shift originating from the environment in $R\mathcal{L}_{\rm I}$. 
The latter term can be taken into account by the redefinition $H_S^\prime=H_S+PH_I$ and
$H_I^\prime=(1-P)H_I$. 
The irreversibility is given by the second (time-retarded)
term. The integral kernel in eq.~(\ref{eq:nz2}) still consists of all powers in $\mathcal{L}_I$
and the dynamics of the reduced density operator $\rho$ of the relevant system depends on
its own whole history. To overcome these difficulties in practically solving eq.~(\ref{eq:nz2}),
one has to make approximations. We begin by assuming that the system bath interaction is weak and
restrict ourselves to the Born approximation, second order in $\mathcal{L}_I$. This allows us 
to replace $\mathcal{L}_{total}$ by $\mathcal{L}_{sys}+\mathcal{L}_{res}$ in the exponent. The
resulting equation is still nonlocal in time. As it is convolutive, it can in principle be solved without
further approximations \cite{Loss03}. To proceed to the more convenient Bloch-Redfield limit, 
we remove the memory firstly by propagating $\rho(t^\prime)$ forward to $\rho(t)$. In principle,
this would require solving the whole equation first and not be helpful. In our case, however, we can observe that the other term in the integral --- the kernel of the equation --- is essentially a bath correlation
function which only contributes at $t-t^\prime<\tau_c$. Using the motional narrowing condition
eq.\ \ref{eq:narrow}, we see that the system is unlikely to interact with the environment in that period
and we can replace the evolution of $\rho$ with the free eveloution, $\rho(t^\prime)=e^{\mathcal{L}_{\rm sys}(t-t^\prime)}\rho(t)$. After this step, the equation is local in time, but the coefficients are still
time-dependent. Now we flip the integration variable $t^\prime\rightarrow t-t^\prime$ and then use the
motional narrowing condition again to send the upper limit of the integral to infinity, realizing 
that at such large time differences the kernel will hardly contribute anyway. We end up with
the Bloch-Redfield equation 
\begin{equation}
\dot{\rho}(t) = P(\mathcal{L}_{\rm sys}+\mathcal{L}_I)\rho(t) + \int_0^\infty\limits dt' P \mathcal{L}_I e^{(1-P)(\mathcal{L}_{\rm sys}+\mathcal{L}_{\rm res})t'} (1 - P)\mathcal{L}_I \rho(t) .
\end{equation}
The Bloch-Redfield equation is of Markovian form, however, by properly
using the free time evolution of the system (back-propagation), they take into
account all bath correlations which are relevant within the Born approximation 
\cite{Hartmann00}. In \cite{Hartmann00}, it has also been shown 
that in the bosonic case the Bloch-Redfield theory is numerically equivalent 
to the path-integral method.

The resulting Bloch-Redfield equations for the reduced density matrix $\rho$ in the eigenstate 
basis of $H_{\rm sys}$ then read
\cite{Weiss99}
\begin{equation}
\dot{\rho}_{nm}(t) = -i\omega_{nm}\rho_{nm}(t)-\sum_{k,\ell} R_{nmk\ell}\rho_{k\ell}(t)
\label{redfield} \ ,
\end{equation}
where $R_{nmk\ell}$ are the elements of the Redfield tensor and the $\rho_{nm}$
are the elements of the reduced density matrix.

The Redfield tensor has the form \cite{Weiss99,Blum96}
\begin{equation}
R_{nmk\ell}=\delta_{\ell m} \sum_{r} \Gamma^{(+)}_{nrrk} + \delta_{nk} \sum_{r} \Gamma^{(-)}_{\ell rrm} - \Gamma^{(+)}_{\ell mnk} - \Gamma^{(-)}_{\ell mnk}. \label{redfield_ten}
\end{equation}

The rates entering the Redfield tensor elements are given
by the following Golden-Rule expressions \cite{Weiss99,Blum96}
\begin{eqnarray}
\Gamma^{(+)}_{\ell mnk} & = & \hbar^{-2} \int_{0}^{\infty} \limits dt \ e^{-i\omega_{nk}t}
\langle \tilde{H}_{I,\ell m}(t)\tilde{H}_{I,nk}(0) \rangle 
\label{eq:plusrate}\\
\Gamma^{(-)}_{\ell mnk} & = & \hbar^{-2} \int_{0}^{\infty} \limits dt \ e^{-i\omega_{\ell m}t}
\langle \tilde{H}_{I,\ell m}(0)\tilde{H}_{I,nk}(t) \rangle \ ,
\label{eq:minusrate}
\end{eqnarray}
where $H_I$ appears in the interaction representation
\begin{equation} 
\tilde{H}_{I}(t) = \exp(iH_{\rm res}t/\hbar)\ H_I\ \exp(-iH_{\rm res}t/\hbar) 
\label{eq:ww}.
\end{equation}
$\omega_{nk}$ is defined
as $\omega_{nk}=(E_n-E_k)/\hbar$. In a two-state system, the coefficients $\ell$, $m$, $n$ and $k$ 
stand for either $+$ or $-$ representing
the upper and lower eigenstates. The possible values of $\omega_{nk}$ in a TSS are
$\omega_{++}=\omega_{--}=0$, $\omega_{+-}=\frac{2\delta}{\hbar}$ and
$\omega_{-+}=-\frac{E}{\hbar}$, where $E$ is the energy splitting between
the two charge eigenstates with $E=\sqrt{\epsilon^2 + \Delta^2}$. Now we apply the secular approximation, which again refers to weak damping, to discard many rates in the Redfield
tensor as irrelevant. 
The details of this approximation are most transparent in the multi-level case and will be discussed
in more detail in section \ref{ch:secular}. In the TSS case, the secular approximation holds whenever
the Born approximation holds. After the secular approximation, the Bloch-Redfield equation coincides
with the Bloch equation with
\begin{eqnarray}
1/T_1 & = & \sum_n\limits R_{nnnn} = R_{++++} + R_{----} =\Gamma_{-++-}+\Gamma_{-++-}\\
1/T_2 & = & {\rm Re} (R_{nmnm}) = {\rm Re} (R_{+-+-}) = {\rm Re} (R_{-+-+}) \nonumber\\
&=&{\rm Re} (\Gamma_{+--+}+\Gamma_{-++-}+\Gamma_{----}+\Gamma_{++++}-\Gamma_{--++}-\Gamma_{++--}).\nonumber\\
&=&\frac{1}{2T_1}+\frac{1}{T_\phi} \label{eq:ttwo}
\end{eqnarray}
Here, we have introduced the dephasing rate $T_\phi^{-1}$. 
The relaxation rate is given by the time evolution of the {\it diagonal} elements,
and the dephasing rate by the {\it off-diagonal} elements of the reduced density matrix $\rho$.

The factor of two in the formula connecting $1/T_2$ and $1/T_1$ appears to 
be counterintuitive, as we would expect that energy relaxation definitely also leads to dephasing,
without additional factors. This physical picture is also correct, but one has to take into account that
there are {\em two} channels for dephasing --- clockwise and counterclockwise precession --- which
need to be added. In fact, this is the reason why the same factor of two appears in the positivity 
condition for the density matrix, see section \ref{ch:bloch}. Another view is to interpret the diagonal matrix elements 
as  classical probabilities, the absolute square of a eigenfunctions of the Hamiltionian, $|\psi_1|^2$, whereas the off-diagonal terms constitute amplitudes, $\psi_2^\ast \psi_1$. Being squares, probabilities decay twice as fast as amplitudes. This point will be discussed further later on in the 
context of multi-level decoherence, eq.\ \ref{eq:positivity}. 

The imaginary part of the Redfield tensor elements that are relevant for the dephasing rate
$\Im (R_{+-+-})$  provides a renormalization of the coherent oscillation frequency
$\omega_{+-}$, $\delta\omega_{+-}=\Im(\Gamma_{+--+}+\Gamma_{-++-})$. If the renormalization of the oscillation frequency gets larger than
the oscillation frequency itself, the Bloch-Redfield approach with its weak-coupling approximations does
not work anymore. By this, we have a direct criterion for the validity of the calculation. 

Finally, the stationary population is given by
\begin{equation}
S_{z,eq}=\frac{\Gamma_{-++-}-\Gamma_{+--+}}{\Gamma_{-++-}+\Gamma_{+--+}}=\tanh\left(\frac{\hbar\omega_{+-}}{2k_B T}\right)
\end{equation}
where in the last step we have used the property of detailed balance
\begin{equation}
\Gamma_{nmmn}=\Gamma_{mnnm}e^{-\omega_{mn}/k_BT}
\end{equation}
which holds for any heat bath in thermal equilibrium and is derived e.g.\ in References \cite{Weiss99,Ingold98,Callen51}.
 
A different kind of derivation with the help of Keldysh diagrams 
for the specific case of an single-electron transistor (SET)
can be found in the Appendix of Ref.~\cite{Makhlin01}.

Very recent results \cite{Gutmann05,Thorwart05} confirm  that without the secular approximation, 
Bloch-Redfield theory preserves complete positivity only in the pure
dephasing case (with vanishing coupling $\Delta = 0$ between the qubit states). In all other cases, complete positivity is violated at short time scales. Thus only in the pure dephasing regime is 
the Markovian master equation
of Lindblad form \cite{Lindblad76} as typically postulated in mathematical physics. In all other 
cases the Lindblad theorem does {\em not} apply. This is not an argument against Bloch-Redfield --- the 
Markovian shape has been obtained as an approximation which coarse-grains time, i.e.
it is not supposed to be valid on  short time intervals. Rather one has to question the generality
of the Markov approximation \cite{Lidar04} at low temperature. Note,\ that in some cases the violation of positivity 
persists and one has to resort to more elaborate tools for consistent results \cite{Thorwart05}

\subsubsection{Rates for the Spin-Boson model and their physical meaning \label{sbredfield}}

This technique is readily applied to the spin boson Hamiltonian eq. (\ref{eq:hamiltonian_tss}). The structure of the golden rule rates eqs. (\ref{eq:plusrate} and \ref{eq:minusrate}) become rather transparent --- the
matrix elements of the interaction taken in the energy eigenbasis measure symmetries and selection rules whereas the time integral essentially leads to energy conservation. 

In particular, we can identify the energy relaxation rate
\begin{equation}
\frac{1}{T_1}=\frac{\Delta^2}{E^2}S(E).
\label{eq:teins}
\end{equation}
The interpretation of this rate is straightforward --- the system has to make a transition, exchanging energy $E$ with the 
environment using a single Boson. The factor $S(E)=J(E)(n(E)+1+n(E))$ captures the density of Boson 
states $J(E)$ and the sum of the rates for emission proportional to $n(E)+1$ and absorption proportional to $n(E)$ of 
a Boson. Here, $n(E)$ is the Bose function. The prefactor is the squared cosine of the angle between the coupling to the noise and the qubit Hamiltonian, i.e. it is maximum if --- in the basis of qubit eigenstates --- the bath couples to the qubit in a fully off-diagonal way. This is reminiscent of the standard
square of the transition matrix element in Fermi's golden rule. 

The flip-less contribution to $T_2$ reads
\begin{equation}
\frac{1}{T_\phi}=\frac{\epsilon^2}{2E^2}S(0).
\label{eq:tzwei}
\end{equation}
It accounts for the dephasing processes which do not involve a transition of the qubit. Hence, they exchange zero energy with the environment and $S(0)$ enters. The prefactor measures which fraction
of the total environmental noise leads to fluctuations of the energy splitting, i.e., it is complemetary 
to the transition  matrix element in $T_1$ ---  the component of the noise {\em diagonal} in the basis of energy eigenstates leads to pure dephasing. The zero frequency argument is a consequence of the Markov approximation. More physically, it can be
understood as a limiting procedure involving the duration of the experiment, which converges 
to $S(0)$ under the motional narrowing condition. Details of this procedure and its limitations
will be discussed in the next section.

Finally, the energy shift
\begin{equation}
\delta E=\frac{\Delta^2}{E^2}{\cal P}\int d\omega \frac{J(\omega)}{E^2-\omega^2},
\label{eq:lamb}
\end{equation}
where ${\cal P}$ denotes the Cauchy mean value, is analogous to the energy shift in second order 
perturbation theory, which collects all processes in which a virtual Boson is emitted and reabsorbed, i.e. 
no trace is left in the environment. Again, the prefactor ensures that the qubit makes a virtual transition
during these processes. For the Ohmic case, we find
\begin{equation}
\delta E=\alpha E \frac{\Delta^2}{E^2}\log \left(\frac{\omega_c}{E}\right)
\end{equation}
provided that $\omega_c\gg E$. Thus, the energy shift explicitly depends on the ultraviolet cutoff. In fact,
$\delta E\simeq E$ would be an indicator for the breakdown of the Born approximation. Thus, we can 
identify two criteria for the validity of this approximation, $\alpha\ll1$ and $\alpha\log(\omega_c/E)\ll1$. The latter is more confining, i.e. even if the first one is satisfied, the latter one can be violated. Note that in some parts of the open quantum systems literature, the justification and introduction of this ultraviolet cutoff is discussed extensively. The spectral densities we have computed so far in the previous sections have always had an
intrinsic ultraviolett cutoff, e.g. the pure reactive response of electromagnetic circuits at high frequencies.

\subsection{Engineering decoherence}

The picture of decoherence we have at the moment apparently allows to engineer the decoherence properties --- which we initially percieved as something deep and fundamental --- using a limited set of formulae, eqs. \ref{eq:teins}, \ref{eq:tzwei} and \ref{eq:jthroughz}, see Refs. \cite{EPJB03,Makhlin01}
these equations
have been applied to designing the circuitry around quantum bits. This is, however, not the end of the story. After this process had been mastered to sufficient degree, decoherence turned out to be limited by more intrinsic phenomena, and by phenomena not satisfactorily described by the Bloch-Redfield technique. This will be the topic of the next section.

\section{Beyond Bloch-Redfield\label{ch:beyond}}

It is quite surprising that a theory such as Bloch-Redfield, which contains a Markov approximation, works so well at the low temperatures at which superconducting qubits are operated,
even though correlation functions at low temperatures decay very slowly and can have significant power-law time tails. The main reason for this is the motional narrowing condition mentioned above, which essentially states that a very severe Born approximation, making the system-bath interaction the lowest energy/longest time in the system, will also satisfy that condition. This is analogous to the textbook derivation of Fermi's golden rule \cite{Cohen92,Sakurai67}, where the perturbative interaction is supposed to be the slowest process involved. In this section, we are going to outline the limitations of this approach by comparing to practical alternatives. 

Before proceeding we would also like to briefly comment
on the general problem of characterizing the environment
in an open quantum system. The most general environment
is usually assumed to induce a completely positive linear
map (or "quantum operation") on the reduced density matrix.
The most general form of such a map is known as the Krauss
operator-sum representation, although such a representation
is not unique, even for a given microscopic system-bath model
like the one considered here. A continuous-time master
equation equivalent to a given Krauss map is provided by the
Lindblad equation, but the form of the Lindblad equation is
again not unique. The Lindblad equation gives the most
general form of an equation of motion for the reduced density
matrix that assures complete positivity and conserves the
trace; however, the Marvok and Born approximations are often
needed to construct the specific Lindblad equation corresponding
to a given microscopic model. The Markov approximation is a
further additional simplification, rendering the dynamics to that
of a semigroup. A semigroup lacks an inverse, in accordance
with the underlying time-irreversibility of an open system.
However, like the unitary group dynamics of a closed system,
the semigroup elements can be generated by exponentials of
non-Hermitian "Hamiltonians", greatly simplifying the analysis.
The Bloch-Redfield master equation also has a form similar to
that of the Lindblad equation, but there is one important
difference: Bloch-Redfield equation does not satisfy
complete positivity for all values of the diagonal and
off-diagonal relaxation parameters. If these parameters
are calculated microscopically (or are obtained empirically),
then complete positivity will automatically be satisfied, and the
Bloch-Redfield equation will be equivalent to the Lindblad
equation. Otherwise inequalities have to be satisfied by
the parameters in order to guarantee complete positivity.

\subsection{Pure dephasing and the independent Boson model}

We start from the special case $\Delta=0$ of the spin-Boson model, also known as the independent Boson model \cite{Mahan00}. We will discuss, how this special case can be solved exactly for a variety of initial conditions. Restricting the analysis to this case is a loss of generality. In particular, as the qubit
part of the Hamiltonian commutes with the system-bath coupling, it cannot induce transitions between
the qubit eigenstates. Thus $1/T_1=0$ to all orders as confirmed by eq.\ \ref{eq:teins} and
$1/T_2=S(0)$ following eq.\ \ref{eq:tzwei}.
Still, it allows to gain insight into a number of phenomena and the validy of the standard approximations. 
Moreover, the results of this section have been confirmed based on a perturbative diagonalization scheme valid for gap or super-ohmic environmental 
spectra \cite{Visibility}.

\subsubsection{Exact propagator}

As the qubit and the qubit-bath coupling commute, we can construct the exact propagator of the system. 
We go into the interaction picture. The system-bath coupling Hamiltonian then reads
\begin{equation}
H_{SB}(t)=\frac{1}{2}\sigma_z\sum_j \lambda_j(a_ie^{-i\omega_j t}+a_i^\dagger e^{i\omega_j t}).
\end{equation}
The commutator of this Hamiltonian with itself taken at a different time is a c-number. Consequently,
up to an irrelevant global phase, we can drop the time-ordering operator ${\mathcal T}$ in the
propagator \cite{Sakurai67,Mahan00} and find
\begin{eqnarray}
U(t,t^\prime)&=&{\mathcal T} \exp\left(-\frac{i}{\hbar}\int_{t^\prime}^t dt^\prime H_{SB} (t^\prime)\right)\\
&=&\exp\left(\sigma_z\sum_i \frac{\lambda_i}{2\hbar\omega_i}\left(a_i^\dagger \left(e^{i\omega_i (t-t^\prime)-1}\right)-a_i \left(e^{-i\omega_i (t-t^\prime)-1}\right)\right)\right).\nonumber
\end{eqnarray}
In order to work with this propagator, it is helpful to reexpress it using shift operators
$D_i(\alpha_i)=\exp(\alpha a^\dagger-\alpha^\ast a)$ as
\begin{equation}
U(t,t^\prime)=\prod_j D_j\left(\sigma_z\frac{\lambda_j}{2\hbar\omega_j}\left(e^{i\omega_j (t-t^\prime)}-1\right)\right).
\label{eq:propagator}
\end{equation}
This propagator can be readily used to compute observables. The main technical step remains to 
trace over the bath using an appropriate initial state. 
The standard choice, also used for the derivation of the Bloch-Redfield equation, is the 
factorized initial condition with the bath in thermal equilbrium, i.e. the initial density matrix
\begin{equation}
\rho(0)=\rho_q \otimes e^{-H_B/kT}
\end{equation}
where we use the partition function $Z$ \cite{Landau84}. The expectation 
value of the displacement operator between number states is $\left\langle n \right | D( \alpha ) \left | n \right\rangle=e^{-(2n+1)\left | \alpha \right |^2/2}$. We start in an arbitrary pure initial state of the qubit
\begin{equation}
\rho_q=\left | \psi \right\rangle\left\langle \psi \right |, \quad \left | \psi \right\rangle = \cos\frac{\theta}{2} \left | 0\right\rangle + \sin\frac{\theta}{2} e^{i\phi} \left | 1\right\rangle.
\end{equation}
Using these two expressions, we can compute the exact reduced density matrix, expressed through the three spin projections
\begin{eqnarray}
\left\langle \sigma_x\right\rangle (t)&=&\sin\theta\cos (Et+\phi)e^{-K_f(t)}\label{eq:sigmaxdecay}\\
\left\langle \sigma_y\right\rangle (t)&=&\sin\theta\sin (Et+\phi)e^{-K_f(t)}\\
\left\langle \sigma_z\right\rangle (t)&=&\cos\theta
\end{eqnarray}
where we have introduced the exponent of the envelope for factorized initial conditions, 
\begin{equation}
K_f(t)=\int \frac{d\omega}{\omega^2}S(\omega)(1-\cos\omega t)
\end{equation}
which coincides with the second temporal integral of  the semiclassical correlation function
$S(t)$, see eq.\ \ref{eq:somega}. What does this expression show to us? At short times, we always have $K_f(t)\propto \frac{t^2}{2}\int d\omega S(\omega)$, which is an integral dominated by large frequencies and thus usually depends on the cutoff of $S(\omega)$.  At long times, it is instructive to rewrite this as
\begin{equation}
K_f(t)=t\int d\omega \delta_\omega (t) S(\omega)
\end{equation}
where we have introduced $\delta_\omega (t)=2\frac{\sin^2\omega t/2}{\omega^2 t}$, which approaches $\delta(\omega)$ as $t\longrightarrow\infty$. Performing this limit more carefully, we can do an
asymptotic long-time expansion.  Long refers to the internal time scales of the noise, i.e. the reciprocal of the internal frequency scales of $S(\omega)$, including $\hbar/kT$, 
$\omega_c^{-1}$. The expansion reads
\begin{equation}
K_f(t)=-t/T_2+\log v_F+O(1/t)
\label{eq:longtime}
\end{equation}
with $1/T_2=S(0)$ as in the Bloch-Redfield result and $\log v_F=\mathcal{P}\int \frac{d\omega}{\omega^2}S(\omega)$. Here, $\mathcal{P}$ is the Cauchy mean value regularizing the singularity at $\omega=0$. To highlight the meaning of $v_F$, the visibility for factorized initial conditions, we plug this expansion into eq.\ \ref{eq:sigmaxdecay} and see that 
$\left\langle \sigma_x\right\rangle (t)=v_F\sin\theta\cos (Et+\phi)e^{-t/T_2+O(1/t)}$. Thus, a long-time 
observer of the full dynamics sees exponential decay on a time scale $T_2$ which coincides with the Bloch-Redfield result for the pure dephasing situation,  but with an
overall reduction of amplitude by a factor $v<1$. This is an intrisic 
loss of visibility \cite{Vion02,Simmonds04}. Several experiments have reported a loss of visibility, to which this may be a contribution. Note that by improving detection schemes, several other sources of reduced visibility have been eliminated \cite{Lupascu04,Wallraff05}. 

This result allows a critical assessment of the Born-Markov approximation we used in the derivation of 
the Bloch-Redfield equation. It fails to predict the short-time dynamics --- which was to be expected as the
Markov approximation is essentially a long-time limit. In the long time limit, the exponential shape 
of the decay envelope and its time constant are predicted correctly, there are no higher-order corrections to $T_2$ at the pure dephasing point. The value of $T_2$ changes at finite $\Delta$, see Refs. \cite{Leggett87,Weiss99,Visibility}. A further description of those results would however be far beyond the scope of this chapter and can be found in ref \cite{Visibility}. Finally, we can see how both short and long-time dynamics are related: the short-time (non-Markovian) dynamics 
leaves a trace in the long-time limit, namely a drop of visibility. 

We now give examples for this result. In the Ohmic case $J(\omega)=\alpha\omega e^{-\omega/\omega_c}$ at $T=0$. Hence, we can right away compute $\dot{K}_f(t)$ and obtain 
$K_f(t)=\frac{\alpha}{2}\log(1+(\omega_ct)^2)$ by a single time integral.  
In agreement with the formula for $T_2$, see eqs. \ref{eq:ttwo}, \ref{eq:tzwei}, the resulting
decay does not have an exponential component at long time but keeps decaying as a power law, 
indicating vanishing visibility. 

At finite temperature, the computation follows the same idea but leads to a more complicated result.
We give the expression from Ref. \cite{Gorlich88} for a general power-law bath $J_q(\omega)=\alpha_q \omega^q\omega_c^{1-q}
e^{-\omega/\omega_c}$, 
\begin{eqnarray}
K_f(t)&=&2{\rm Re}\left\lbrace \alpha_q\Gamma(q-1)\left(1-(1+i\omega_c t)^{1-s}+\left(\frac{\hbar\omega_c}{kT}\right)\times\right.\right.\\
&&\left.\left. \times\left[2\zeta(s-1,\Omega)-\zeta\left(s-1,\Omega+\frac{ikTt}{\hbar}\right)-\zeta\left(q-1,\Omega-\frac{ik\tau kT}{\hbar}\right)\right]\right)\right\rbrace\nonumber
\end{eqnarray}
where we have introduced $\Omega=1+k_BT/\hbar\omega_0$ and the generalized Riemann zeta
function, see \cite{Abramowitz65} for the definition and the mathematical properties
used in this subsection. 
This exact result allows to analyze and quantify the decay envelope by computing the
main parameters of the decay, $v_F$ and $1/T_2$. We will restrict ourselves to the scaling
limit, $\omega_c\gg 1/t, kT$. For the Ohmic case, $q=1$, we obtain at finite temperature  
$1/T_2=2\alpha kT/\hbar$ and $v_F=(kT/\omega_c)^\alpha$. This result is readily understood. The form
of $T_2$ accounts for the fact that an Ohmic model has low-frequency noise which is purely thermal 
in nature. The visibility drops with growing $\omega_c$ indicating that if we keep adding high frequency modes they all contribute to lost visibility. It is less intuitive that $v_F$ drops with lowering the
temperature, as lowering the temperature generally reduces the noise. This has to be discussed together with the $1/T_2$-term, remembering that $1/T_2$ is the leading and $v_F$ only the sub-leading order of the long time expansing eq.\ \ref{eq:longtime}: At very low temperatures, the crossover to the exponential long-time
decay starts later and the contribution of non-exponential short time dynamics gains in relative significance. Indeed, at any given time, the total amplitude gets enhanced by lowering the temperature. 

In order to emphasize these general observations, let us investigate the super-Ohmic case with $q\ge 3$. Such spectral functions can be realized in electronic circuits by RC-series shunts \cite{PRB051}, they also
play a significant role in describing phonons. For $q>3$, the exponential component vanishes,
$1/T_2=0$ and $v_q=\exp [-2\alpha_q\Gamma (q-1)]$. Thus, we obtain a massive loss of visibility but no exponential envelope at all. This highlights the fact that $v$ and $1/T_2$ are to be
considered independent quantifiers of non-Markovian decoherence and that the latter accounts for 
environmental modes of relatively low frequency whereas $v$ is mostly influenced by the fast modes
between the qubit frequency and the cutoff. 

Before outlining an actual microscopic scenario, we generalize the initial conditions of our calculation. 

\subsubsection{Decoherence for non-factorzing initial conditions\label{nofactor}}

Our propagator, eq. \ref{eq:propagator}, is exact and can be applied to any initial density matrix. 
We start from an initial wave function
\begin{equation}
\left|\psi\right\rangle =\left|0\right\rangle {\prod}
_n D(z^0_i/2\lambda_i\hbar\omega_i)\left|0\right\rangle_i+\left|1\right\rangle {\prod}_n D(z^1_i/2\lambda_i\hbar\omega_i)\left|0\right\rangle_i
\label{eq:dressed}
\end{equation}
where we have introduced sets of dimensionless coefficients $z_i^{0/1}$. 
It would be straightforward to introduce $\theta$ and $\phi$, which we will stay away from in order to
keep the notation transparent. The factorized initial condition corrsponds to 
$z_i^{0/1}=0$. 

This structure has been chosen in order to be able to obtain analytical results, using the structure of
the propagator expressed in displacement operators, eq. \ref{eq:propagator} and the multiplication rules for these operators \cite{Walls94}. Note that the choice of coherent states to entangle the qubit with is not a severe restriction. It has
been shown in quantum optics in phase space, that essentially each density matrix of an harmonic
oscillator can be decomposed into coherent states using the Wigner or Glauber P phase 
space representations, see e.g. 
\cite{Schleich01}.  Physically, 
the initial state eq.\ \ref{eq:dressed} 
corresponds to the qubit being in a superposition of two dressed states. Of specific significance is the initial
condition which minimizes the sytem bath-interaction in the Hamiltonian eq.\ \ref{eq:hamiltonian_tss}, nameley $z_i^0=-z_i^{-1}=-1$. 

We can again compute all three spin projections of the qubit. The essence of the decoherence behavior
is captured in the symmetric initial state, $z_i^0=-z_i^1$ for all $i$
\begin{equation} 
\left\langle\sigma\right\rangle_x=\cos Et e^{-K(t)}
\end{equation}
very similar to eq.\ \ref{eq:sigmaxdecay} in the factorized case, but now with 
\begin{eqnarray*}
K(t)&=&-\frac{1}{2}\int_0^\infty \frac{d\omega}{\omega^2}
J(\omega)\left[(u(\omega)+1)^2+v^2(\omega)+1-\right.\\
&&\left.\left.-2\left(1+u(\omega)\cos\omega
t+v(\omega)\sin\omega t\right)\right]\right)\
\end{eqnarray*}
where we have taken a continuum limit replacing the complex numbers $z_i^0$ by the real function $u(\omega)+iv(\omega)$. This form connects to the factorized case by setting $u=v=0$.  For any other 
choice of $u$ and $v$, the initial conditions are entangled. 

We can make a few basic observations using this formula: The initial amplitude $e^{-K(0)}$ is controlled
through 
\begin{equation}
K(0)=\int \frac{d\omega}{2\omega^2}\left[u^2(\omega)+v^2(\omega)\right],
\end{equation}
thus for any initial condition which is more than marginally entangled (meaning that the integral is 
nonzero), the initial amplitude is smaller 
than unity. On the other hand, the time-dependence of $K(t)$ can be completely eliminated by
chosing an initial condition $u=-1,v=0$. This condition 
minimizes the system-bath part of the total energy in the sense of variation with respect to $u$ and
$v$. This choice of initial state also minimizes the total energy if the oscillators are predominantly at
high frequency, whereas for the global minimum one would rather chose a factorized state for the 
low-frequency oscillators. Physically, this corresponds to an optimally dressed state of the qubit surrounded by an oscillator dressing cloud. The overlap of these clouds reduces the amplitude
from the very beginning but stays constant, such that the long-time visibility
\begin{equation}
v_g=\int \frac{d\omega}{2\omega^2}\left[(u(\omega)-1)^2+1+v^2(\omega)\right],
\end{equation}
is maximum. Note that this reduces to the result for $v_F$ for $u=v=0$. 

What can we learn from these results? We appreciate that initial conditions have a significant and
observable effect on the decoherence of a single qubit. The choice of the physically appropriate 
initial condition is rather subtle and depends on the experiment and environment under consideration. A free induction decay experiment as described here does usually not start out of the blue.
It is launched using a sequence of preparation pulses taking the state from a low temperature 
thermal equilibrium to the desired initial polarization of the qubit. Thus, from an initial equilbrium state
(for some convenient setting of the qubit Hamiltonian), 
the fast preparation sequence initiates nonequilibrium correlations thus shaping $u$ and $v$. 
Furthermore, if the interaction to the environment is tunable such as in the case of the
detectors discussed previously in section \ref{jcircuit3}, the initial condition interpolates between factorized (rapid switching
of the qubit-detector coupling) and equilibrium (adiabatic switching). 

At this point, we can draw conclusions about the microscopic mechanism of the loss of visibility
and other short-time decoherence dynamics. The picture is rooted on the observation that the ground state of the coupled system is a dressed state. On the one hand, as described 
above, the overlap of the dressing clouds reduces the final visibility. On the other hand, for nonequilibrium initial conditions such as the factorized one, there is extra energy stored in the system 
compared to the dressed ground state. This energy gets redistributed while the dressing cloud is forming, making it possible for an excitation in the environment with an 
extra energy $\delta E$ to be created leading to a virtual intermediate state, followed by another excitation
relaxing, thus releasing the energy $\delta E$ again. It is crucial that this is {\em another} excitation as only processes which leave a trace in the environment lead to qubit dephasing.  Higher-order processes
creating and relaxing the {\em same} virtual excitation only lead to renormalization effects such as 
the Lamb shift, see eq. \ref{eq:lamb}. This explains why the loss of visibility is minimal for dressed initial conditions, where
no surplus excitations are present. 

The Bloch-Redfield technique is a simple and versatile tool which makes good predictions of decoherence rates at low damping. At higher damping,
these rates are mostly joined by renormalization effects extending the Lamb shift in eq.\ \ref{eq:lamb}, see Refs. \cite{Leggett87,Weiss99,Visibility}. However, there is more to decoherence than 
a rate for accurate predictions of coherence amplitudes as a function of time, one has
to take the non-exponential effects into account and go beyond Bloch-Redfield. Other approaches 
can be applied to this system such as rigorous (Born but not Markov) perturbation theory \cite{Loss03}, path-integral techniques  \cite{Leggett87}, \cite{Weiss99}, and renormalization schemes \cite{Kehrein98}.

Note, that these conclusions all address free induction decay. There is little indication on the quality of the Bloch-Redfield theory in the presence of pulsed driving. 

\subsubsection{1/f noise \label{onef}}

In the previous sections we have explored options how to engineer decoherence by influencing 
the spectral function $J(\omega)$ e.g. working with the electromagnetic environment. This has helped
to optimize supercondcuting qubit setups to a great deal, down to the level where the noise intrinsic to
the material plays a role. In superconductors, electronic excitations are gapped \cite{Tinkham96} and the electron 
phonon interaction is weak due to the inversion symmetry of the underlying crystal everywhere except
close to the junctions \cite{Ioffe04}. The most prominent
source of intrinsic decoherence is thus $1/f$ noise. $1/f$ noise - noise whose spectral function 
behaves following $S(\omega)\propto 1/\omega$, is ubiquitous in solid-state systems. This spectrum 
is very special as all the integrals in our discussion up to now would diverge for that spectrum. $1/f$ 
typically occurs due to slowly moving defects in strongly disordered materials. In Josephson devices,
there is strong evidence for $1/f$ noise of gate charge, magnetic flux, and critical current, leading to a
variety of noise coupling operators (see Ref. \cite{Harlingen04} for an overview). 
Even though
there does not appear to be a fully universal origin, a "standard" model of $1/f$ noise has been identified \cite{Dutta81,Weissman88}: The fundamental unit are two-state fluctuators, i.e. two state systems which couple to the device under consideration and which couple to an external heat bath making them jump between two positions. The switching process consists of uncorrelated switching events, i.e. the distribution of times between these switches is Poissoinan. If we label the  mean time between switches as $\tau$, the spectral function of this process is $S_{RTN}=S_0\frac{1/\tau}{1+\tau^2\omega^2}$. This phenomenon alone is called random telegraph noise (RTN). Superimposing such fluctuators with a flat distribution of switching
times leads to a total noise spectrum proportional to $1/f$. Nevertheless, the model stays different from an 
oscillator bath. The underlying thermodynamic limit is usually not reached as it is approached more
slowly: Even a few fluctuators resemble $1/f$ noise within the accuracy of a direct noise measurement.  Moreover, as  we are interested in very small devices such as qubits, only a few fluctuators are effective and experiments can often resolve them directly \cite{Wakai87}. Another way to see this is to realize that the RTN spectrum is highly non-Gaussian: A two - state distribution can simply not be fitted by a single Gaussian, all its higher cumulants of distribution are relevant.  This non-Gaussian component only vanishes slowly when we increase the system size and is significant for the case of qubits.  

A number of studies of models taking this aspect into account have been published \cite{Paladino02,Grishin05,Shnirman05,Galperin03,Faoro05,PRL05}. A highly simplified
version is to still take the Gaussian assumption but realize that there is always a slowest fluctuator, thus
the integrals in $K_f(t)$ can be cut off at some frequency $\omega_{IR}$ at the infrared (low frequency) end of the spectrum, i.e. using 
the spectral function
\begin{equation}
S(\omega)=\frac{E_{1/f}^2}{\omega}\theta(\omega-\omega_{IR})
\end{equation}
with $\theta$ the Heaviside unit step function, we approximately find \cite{Cottet02,Martinis03,Shnirman02}
\begin{equation}
e^{-K(t)}\simeq (\omega_{IR}t)^{-(E_{1/f}t/\pi\hbar^2)^2}
\end{equation}
so we find the Gaussian decay typical for short times - short on the scale of the correlation time of the
environment, which is long as the spectrum is dominated by low frequencies - with a logarithmic 
correction. 

At the moment, forefront research works at understanding more detailed models of $1/f$ noise and
understand the connection between the strong dephasing and a possible related relaxation mechanism 
at high frequencies. On the other hand, experiments work with materials to avoid $1/f$-noise at its source. Generally, slow noise up to a certain level can be tolerated using refocusing techniques
such as simple echo or the Carr-Purcell-Gill-Meiboom pulse sequence \cite{PRAR05,Carr54,Faoro04,Falci04,Shiokawa04,Bertet05b}, the power and potential 
of which has been
demonstrated both experimentally and theoretically. 

\section{Decoherence in coupled qubits}

To conclude, we want to outline how to go beyond a single to multiple qubits and identifty the underlying challenges. On that level, much less is known both theoretically and experimentally. The variety of 
physically relevant Hamiltonians is larger. One extreme case is fully uncorrelated noise, e.g. originating
from effects in the junctions or qubit-specific Hamiltonians, 
\begin{equation}
H=H_{Q1}+H_{Q2}+H_{QQ}+H_{Q1B1}+H_{B1}+H_{Q2B2}+H_{B2}
\label{eq:getrenntebaeder}
\end{equation}
this is simply the sum of two single-qubit decoherence Hamiltonians in distinct Hilbert spaces, consisting 
of qubit Hamiltonians $H_{Qi}$, $i=1,2$, baths $H_{Bi}$, qubit-bath interaction $H_{QiBi}$ all interacting 
via a qubit-qubit interaction $H_{QQ}$ alone. The other extreme case is collective noise, e.g. long-wavelength
ambient fluctuations or noise shared control lines. This is described by
\begin{equation}
H=H_{Q1}+H_{Q2}+H_{QQ}+H_{Q1B}+H_{Q2B}+H_{B}
\label{eq:corrnoise}
\end{equation}
where both qubits talk to a single bath. The distinction of baths may seem artifical, as this is a special
case of Hamiltonian \ref{eq:getrenntebaeder}: What we 
really mean is that in the interaction picture there is a significant correlation between baths
$\left\langle H_{Q1B}(t) H_{Q2B}(t^\prime)\right\rangle\not=0$. Note, that intermediate cases between
these, a partially correlated model \cite{PRB052}, can be identified in the context of quantum dots.

\subsubsection{The secular approximation\label{ch:secular}}

What does it take to study decoherence here or in other multilevel systems? Basically we can follow all the steps through the derivation of the Bloch-Redfield equation given in section \ref{ch:redfield} up to eq. \ref{eq:ww} until we solve the equation. There, we have already mentioned the secular approximation without explaining its details. 

The essence of the secular approximation is the separation
of time scales. We go back to the interaction representation of equation \ref{redfield}, leading to
\begin{equation}
\dot{\rho}_{nm}^I=\sum_{kl}R_{nmkl}e^{i(\omega_{nm}-\omega_{kl})t}\rho_{kl}^I.
\end{equation}
As the Bloch-Redfield equation is based on a Born approximation, we can expect $|R_{ijkl}|\ll \omega_{mn}$ for all coefficients $i,j,k,l,m,n$ with $m\not=n$. 

In the {\em secular limit}, this also holds true for most frequency splittings 
\begin{equation}
|\omega_{nm}-\omega_{kl}|\gg |R_{nmkl}|
\label{eq:secular}
\end{equation}
besides the inevitable exceptions of $n=m,\, k=l$, and $n=k$ and $m=l$. Whenever condition eq.\ \ref{eq:secular} is satisfied,  the time evolution induced by $R_{nmkl}$ is certainly 
slower than the precession with $\omega_{nm}-\omega_{kl}$ and averages out quickly, hence it can 
be dropped. So the only remaining rates are the cases just mentioned:

For $n=k, m=l$, we have to keep  $R_{nmnm}$. This rate is the dephasing rate for the transition between levels $nm$, see eq. \ref{eq:tzwei}. These rates depend on the pair
of levels we chose and in general they will all be different for different choices of $n$ and $m$, leading to $N(N-1)/2$ different $T_2$-rates for
an $N$-level system. 

For  $n=m$ and $k=l$. The set of these terms splits
off from the rest of the equation, i.e. the diagonal terms of the density matrix (in the eigenstate basis)
decay independent from the off-diagonal terms and obey the following set of equations
\begin{equation}
\dot{P}_n=\sum_n (P_m\Gamma_{m\rightarrow n}-P_n \Gamma_{n\rightarrow m})
\label{eq:pauli}
\end{equation}
which is analogous to the Pauli master equation for classical probabilities. We have identified
$P_n=\rho_{nn}$, the classical probability and the transition rates $\Gamma_{n\rightarrow m}=R_{nnmm}$. Equation \ref{eq:teins} can be solved by Laplace transform, where it reduced to a matrix inversion. This leads to $N$ different independent
energy relaxation channels whose rates are the eigenvalues of the matrix form of the right hand side of eq.\ \ref{eq:pauli}. One of these eigenvalues is always zero representing stable thermodynamic
equilbrium which does not decay. In the two-state case, this leads us to one nonvanishing $T_1$ rate 
representing the only nonzero eigenvalue, given by eq.\ \ref{eq:teins}. The rates generally obey the positivity constraint
\begin{equation}
\sum_n R_{nnnn}\le 2 \sum_{n\not=m} R_{nmnm}
\label{eq:positivity}
\end{equation}
the left hand side being the trace of the relaxation matrix, i.e. the sum over all $T_1$-type rates and
the ricght hand side being the sum over dephasing rates. 
This reduces to $T_2\le 2T_1$ in a two-state system.

In the opposite case, the case  of a approximate {\em Liouvillian degeneracy}, we find a pair of frequencies $|\omega_{nm}-\omega_{kl}|\ll |R_{nmkl}|$ which do not obey the conditions mentioned in the previous
paragraph, such that the secular approximation does not apply to this set of levels and
$R_{ijkl}$ must be kept. In that case, the Bloch-Redfield equation can still be diagonalized numerically,
identifying the relvant modes of decay \cite{vanKampen97}. Note, that Liouvillian degeneracies can appear in non-degenerate systems, promintently the single harmonic oscillator. One practical example for this issue is intermediate-temperature cavity QED \cite{Rau04}

These concepts already found some application in the theoretical literature. We just mention the main 
results here. After the pioneering work \cite{Governale01}, it was realized that the high number of rates 
makes the results difficult to analyze and the performance of quantum gates should be analyzed 
directly \cite{Thorwart02,PRA03,ASSP03}. A key result is that (only) the correlated noise model, eq.\ \ref{eq:corrnoise} permits to use symmetries and encoding into decoherence free subspaces to protect coherence \cite{PRA03,ASSP03,PRB053}, where deviations from perfect symmetry are of relatively low impact
\cite{PRB054}. 

\section{Summary}

In summary, we have provided an introduction to standard methods in decoherence theory as they are 
applied to superconducting qubits. Many of the tools and results are more general and can be applied to
other damped two-state-systems as well. We see that parts of the theory of decoherence --- in particular the part on electromagnetic environments and Bloch-Redfield-Theory --- are really well established by now, only opening the view on more subtle problems connected to memory effects and the interplay
of decoherence and control. 

\section{Acknowledgements}

This work is based on numerous discussions, too many to list. It is based on our own process of learning and explaining together with other group members, such as M. Goorden, H. Gutmann, A. K\"ack, A. Holzner, K. J\"ahne, I. Serban, and J. Ferber. Very importantly, we thank G. Johannson, M. Governale, M. Grifoni, U. Weiss, G. Falci, P. H\"anggi, S. Kohler, P. Stamp, L. Tian, S. Lloyd, and H. Gutmann on the theoretical 
as well as C.H. van der Wal, C.J.P.M. Harmans, J.E. Mooij, T.P. Orlando, J. Clarke, B.L.T. Plourde, and T.L. Robertson on the experimental side. We are very grateful to M. Flatte and I. Tifrea for organizing the NATO-ASI from which this work originates, and NATO for sponsoring it. Also, the questions of both the
participants of the ASI as well as the participants of the course ``T VI: Physics of quantum computing" at LMU were very important in identifying the issues asking for explanation. We are deeply indebted to A.G. Fowler for his careful reading of the manuscript and many suggestions helping to make it at least close to pedagogical. This work was supported by the DFG through Sonderforschungsbereich 631, by ARDA and NSA through ARO grant P-43385-PH-QC, and DAAD-NSF travel grants. 

\bibliographystyle{nato}
%\bibliography{frankslibrary,frankspapers}

\theendnotes

%\printindex
\end{document}